\documentclass{emulateapj}

\usepackage{natbib}

\usepackage{graphicx}
\usepackage{latexsym}
\usepackage{amsfonts,amsmath,amssymb}
\usepackage{url}
\usepackage[utf8]{inputenc}

\shorttitle{Strong Lens Time Delay Challenge: I. Experimental Design}
\shortauthors{Dobler et al.}

\begin{document}

\title{Strong Lens Time Delay Challenge: I. Experimental Design}

\author{Gregory~Dobler\altaffilmark{1,2}}
\author{Christopher~D.~Fassnacht\altaffilmark{3}}
\author{Tommaso~Treu\altaffilmark{4}$^{*}$}
\author{Phil~Marshall\altaffilmark{5}}
\author{Kai~Liao\altaffilmark{6,4}}
\author{Alireza~Hojjati\altaffilmark{7,8}}
\author{Eric~Linder\altaffilmark{9}}
\author{Nicholas~Rumbaugh\altaffilmark{3}}

\altaffiltext{1}{Kavli Institute for Theoretical Physics, University of California Santa Barbara, Santa Barbara, CA 93106, USA.}
\altaffiltext{2}{Center for Urban Science + Progress, New York University, Brooklyn, NY 11201, USA.}
\altaffiltext{3}{Dept.\ of Physics, University of California, 1 Shields Ave., Davis, CA 95616, USA}.
\altaffiltext{4}{Dept.\ of Physics, University of California, Santa Barbara, CA 93106, USA.}
\altaffiltext{5}{Kavli Institute for Particle Astrophysics and Cosmology, P.O. Box 20450, MS29, Stanford, CA 94309, USA.}
\altaffiltext{6}{Dept.\ of Astronomy, Beijing Normal University, Beijing 100875, China}
\altaffiltext{7}{Dept.\ of Physics and Astronomy, University of British Columbia, 6224 Agricultural Road, Vancouver, B.C. V6T 1Z1, Canada}
\altaffiltext{8}{Dept.\ of Physics, Simon Fraser University, 8888 University Drive, Burnaby BC, Canada V5A1S6}
\altaffiltext{9}{Lawrence Berkeley National Laboratory and University of California, Berkeley, CA 94720}

\altaffiltext{*}{Department of Physics and Astronomy, University of California, Los Angeles, CA 90095; tt@astro.ucla.edu}


\begin{abstract}
The time delays between point-like images in gravitational lens
systems can be used to measure cosmological parameters. The number of
lenses with measured time delays is growing rapidly; the upcoming
\emph{Large Synoptic Survey Telescope} (LSST) will monitor $\sim10^3$
strongly lensed quasars.  In an effort to assess the present
capabilities of the community to accurately measure the time delays,
and to provide input to dedicated monitoring campaigns and future LSST
cosmology feasibility studies, we have invited the community to take
part in a ``Time Delay Challenge'' (TDC). The challenge is organized
as a set of ``ladders,'' each containing a group of simulated datasets
to be analyzed blindly by participating teams. Each rung on a ladder
consists of a set of realistic mock observed lensed quasar light
curves, with the rungs' datasets increasing in complexity and realism.
The initial challenge described here has two ladders, TDC0 and
TDC1. TDC0 has a small number of datasets, and is designed to be used
as a practice set by the participating teams. The (non-mandatory)
deadline for completion of TDC0 was the TDC1 launch date, December 1,
2013. The TDC1 deadline was July 1 2014. Here we give an overview of
the challenge, we introduce a set of metrics that will be used to
quantify the goodness-of-fit, efficiency, precision, and accuracy of
the algorithms, and we present the results of TDC0. Thirteen teams
participated in TDC0 using 47 different methods. Seven of those teams
qualified for TDC1, which is described in the companion paper II.
\end{abstract}

\keywords{gravitational lensing  --- methods: data analysis}


\section{Introduction}

As light travels to us from a distant source, its path is deflected by
the gravitational fields of intervening matter.  The most dramatic
manifestation of this effect occurs in strong lensing, when light rays
from a single source can take several paths to reach the observer,
causing the appearance of multiple images of the same source.  These
images will typically be magnified in size and thus total brightness
(because surface brightness is conserved in gravitational lensing).
When the source is variable, the images are observed to vary with
delays between them due to the differing path lengths taken by the
light and the gravitational potential that it passes through.  A common
example of such a source in lensing is a quasar, an extremely luminous
active galactic nucleus at cosmological distance.  From the observations
of the image positions, magnifications, and the time delays between the
multiple images we can measure the mass structure of the lens galaxy
itself (on scales $\geq M_{\odot}$) as well as  a characteristic
distance between the source, lens, and observer. This ``time delay
distance'' encodes the cosmic expansion rate, which in turn depends on
the energy density of the various components in the universe, phrased
collectively as the cosmological parameters.

The time delays themselves have been proposed as tools to study massive
substructures within lens galaxies \citep{KeetonAndMoustakas2009}, and
for measuring cosmological parameters, primarily the Hubble constant,
$H_0$ \citep[see, e.g.,][for a recent example]{SuyuEtal2013}, a method
first proposed by \citet{Refsdal1964}. In the future, we aspire to
measure further cosmological parameters (e.g., dark energy) by combining
large samples of measured time delay distances
\citep[e.g.,][]{Linder2011,Paraficz2009}. It is therefore of great interest to develop
to maturity the powers of lensing time delay analysis for probing the dark
universe.

New wide-area imaging surveys that repeatedly scan the sky to gather
time-domain information on variable sources are coming online, while
dedicated follow-up monitoring campaigns are obtaining tens of time
delays (e.g., the COSMOGRAIL
program\footnote{\url{http://www.cosmograil.org}}). This pursuit will
reach a new height when the \emph{Large Synoptic Survey Telescope}
(LSST) enables the first long baseline multi-epoch observational
campaign on $\sim$1000 lensed quasars \citep{LSSTSciBook}.  However,
using the measured LSST light curves to extract time delays for
accurate cosmology will require detailed understanding of how, and how
well, time delays can be reconstructed from data with real world
properties of noise, gaps, and additional systematic variations. For
example, to what accuracy can time delays between the multiple image
intensity patterns be measured from individual doubly- or
quadruply-imaged systems for which the sampling rate and campaign
length are given by LSST?  In order for time delay errors to be small
compared to errors from the gravitational potential, we will need the
precision of time delays on an individual system to be better than
3\%, and those estimates will need to be robust to systematic
error. Simple techniques such as the ``dispersion'' method
\citep{PeltEtal1994,PeltEtal1996} or spline interpolation through the
sparsely sampled data
\citep[e.g.,][]{TewesEtal2013a} yield time delays which {\it may} be
insufficiently accurate for a Stage IV dark energy experiment. More
complex algorithms such as Gaussian Process modeling
\citep[e.g.,][]{TewesEtal2013a,Hojjati+Linder2013} may hold more
promise. None of these methods have been tested on large scale data
sets.

At present, it is unclear whether the baseline ``universal cadence''
LSST sampling frequency of $\sim10$ days in a given filter and $\sim 4$
days on average across all filters \citep{LSSTSciBook,LSSTpaper} will
enable sufficiently accurate time delay measurements, despite the long
campaign length ($\sim10$ years).  While ``follow up'' monitoring
observations to supplement the LSST light curves may not be feasible at
the 1000-lens sample scale, it may be possible to design a survey
strategy that optimizes cadence and monitoring at least for some
fields.  In order to maximize the capability of LSST to probe the
universe through strong lensing time delays, we must understand the
interaction between the time delay estimation algorithms and the
anticipated data properties.  While optimizing the accuracy of LSST time
delays is our long term objective, improving the present-day algorithms
will benefit the current and planned lens monitoring projects as well.
Exploring the impact of cadences and campaign lengths spanning the range
between today's monitoring campaigns and that expected from a baseline
LSST survey will allow us to simultaneously provide input to current
projects as well as the LSST project, whose exact survey strategy is not
yet decided.

The goal of this work then is to enable realistic estimates of feasible
time delay measurement accuracy to be made with LSST.  We will achieve
this via a ``Time Delay Challenge'' (TDC), in which we have invited the
community to participate.  Independent, blind analysis of plausibly
realistic LSST-like light curves will allow the accuracy of current time
series analysis algorithms to be assessed and will lead to simple
cosmographic forecasts for the anticipated LSST dataset. This work can
be seen as a first step towards a full understanding of systematic
uncertainties present in the LSST strong lens dataset and will also
provide valuable insight into the survey strategy needs of both Stage
III and Stage IV time delay lens cosmography programs. Blind analysis,
where the true value of the quantity being reconstructed is not known by
the researchers, is a key tool for robustly testing the analysis
procedure, without biasing the results by continuing to look for errors
until the correct answer is reached.

This paper is organized as follows. In Section~\ref{sec:light_curves} we
describe the simulated data that we have generated for the challenge,
including some of the broad details of observational and physical
effects that may make extracting accurate time delays difficult, without
giving away information that will not be observationally known during or
after the LSST survey.  Then, in Section~\ref{sec:structure}, we describe
the structure of the challenge, how interested groups can access the
mock light curves, and a minimal set of approximate cosmographic
accuracy criteria that we will use to assess their performance.
Section~\ref{sec:summary} concludes with a brief summary.


\section{Light Curves and Simulated Data}
\label{sec:light_curves}

The intensity as a function of time for a variable source is referred to
as its light curve. For lensed sources, the light curves of images
follow the intrinsic variability of the quasar source, but with
individual time delays that are different for each image. Only the
relative time delays between the images are measurable, since the
unlensed quasar itself cannot be observed. Of course, we do not actually
measure a continuous light curve, but rather discrete values of the intensity at
different epochs.  This sampling of the light curves, the
noise in the photometric measurement, and external effects causing
additional variations in the intensity all provide complications to
estimation of the time delays.


\subsection{Basics}
\label{sec:basics}

The history of the measurement of time delays in lens systems can be
broadly split into three phases.  In the first, the majority of the
efforts were aimed at the first known lens system, Q0957+561
\citep{WalshEtal1979}.  This system presented a particularly difficult
situation for time delay measurements, because the variability was
smooth and relatively modest in amplitude, and because the time delay
was long.  This latter point meant that the annual season gaps when the
source could not be observed at optical wavelengths complicated the
analysis much more than they would have for systems with time delays of
significantly less than one year.  The value of the time delay remained
controversial, with adherents of the ``long'' and ``short'' delays
\citep[e.g.,][]{PressEtal1992a,PressEtal1992b,PeltEtal1996} in
disagreement until a sharp event in the light curves resolved the issue
\citep{KundicEtal1995,KundicEtal1997}.

The second phase of time delay measurements began in the mid-1990s, by
which time tens of lens systems were known, and small-scale but
dedicated lens monitoring programs were conducted
\citep{SchechterEtal1997,BurudEtal2002}.  With the larger number of
systems, there were a number of lenses for which the time delays were
more conducive to a focused monitoring program, i.e., systems with
time delays on the order of 10--150~days.  Furthermore, advances in
image processing techniques, notably the image deconvolution method
developed by \citet{MagainEtal1998}, allowed optical monitoring of
systems in which the image separation was small compared to the
seeing. The monitoring programs, conducted at both optical and radio
wavelengths, produced robust time delay measurements
\citep[e.g.,][]{LovellEtal1998,BiggsEtal1999,FassnachtEtal1999,FassnachtEtal2002,BurudEtal2002a,BurudEtal2002b},
even using fairly simple analysis methods such as cross-correlation,
maximum likelihood, or the ``dispersion'' method introduced by
\citet{PeltEtal1994,PeltEtal1996}.

The third and current phase, which
began roughly in the mid-2000s, has involved large and systematic
monitoring programs that have taken advantage of the increasing amount
of time available on 1--2~m class telescopes.  Examples include the
SMARTS program \citep[e.g.,][]{KochanekEtal2006}, the Liverpool
Telescope robotic monitoring program
\citep[e.g.,][]{GoicoecheaEtal2008}, and the COSMOGRAIL program.  These
programs have shown that it is possible to take an industrial-scale
approach to lens monitoring, operating decade-long campaigns
\citep[e.g.,][]{KochanekEtal2006} and producing very good time delays
\citep[e.g.,][]{TewesEtal2013a,EulaersEtal2013,RathnaKumarEtal2013}.
The next phase, which has already begun, will be lens monitoring from
new large-scale surveys that include time-domain information such as the
Dark Energy Survey, PanSTARRS, and LSST.

Measured time delays constrain the time delay distance
\begin{equation}
D_{\Delta t} = \frac{D_{\rm d} D_{\rm s}}{D_{\rm ds}}
\end{equation}
where $D_{\rm d}$ is the angular diameter distance between observer and
lens, $D_{\rm s}$ between observer and source, and $D_{\rm ls}$ between
lens and source. Note that because of spacetime curvature the
lens-source distance is not the difference between the other two. The
time delay distance will be inversely proportional to the Hubble
constant $H_0$, the current cosmic expansion rate that sets the scale of
the universe, but the distances also involve the matter and dark energy
densities, and the dark energy equation of state.

The accuracy of $D_{\Delta t}$ derived from the data for a given lens
system is dependent on both the mass model for that system as well as
the precision measurement of the lensing observables.  Typically,
positions and fluxes (and occasionally shapes if the source is resolved)
of the images can be obtained to sub-percent accuracy (see, e.g., the
COSMOGRAIL results), but time delay precisions are usually on the order
of days, or a few percent, for typical systems \citep[see
e.g.,][]{TewesEtal2013b}.  Measuring time delays to this level has required continuous
monitoring over months to years.  However, wide area surveys are disadvantaged
in this aspect, as they only return
to a given patch of sky every few nights, sources are only visible from
a given point on the Earth for certain months of the year, and bad
weather can lead to data gaps.


\subsection{Simulating light curves}
\label{sec:simulate}

Simulating the observation of a multiply-imaged quasar involves four
conceptual steps:

\begin{enumerate}

\item The quasar's intrinsic light curve in a given optical band is
generated at the accretion disk of the black hole in an active galactic
nucleus (AGN).

\item The foreground lens galaxy causes multiple imaging, leading to two
or four lensed light curves that are offset from the intrinsic light curve
(and each other) in both amplitude (due to magnification), and time.

\item Time dependent amplitude fluctuations due to microlensing by stars
in the lens galaxy are generated \emph{on top of} (and
\emph{independently} for) each light curve.

\item The delayed and microlensed light curves are sparsely, but
simultaneously, ``sampled'' at the observational epochs, with the
measurements adding noise.

\end{enumerate}

In the next sections we describe the simulation of each of these steps
in some detail during the generation of the challenge mock LSST light
curve catalog.


\subsection{Intrinsic AGN Light Curve Generation}
\label{sec:car}

\begin{figure*}[!ht]
\begin{center}
\includegraphics[width=0.75\textwidth]{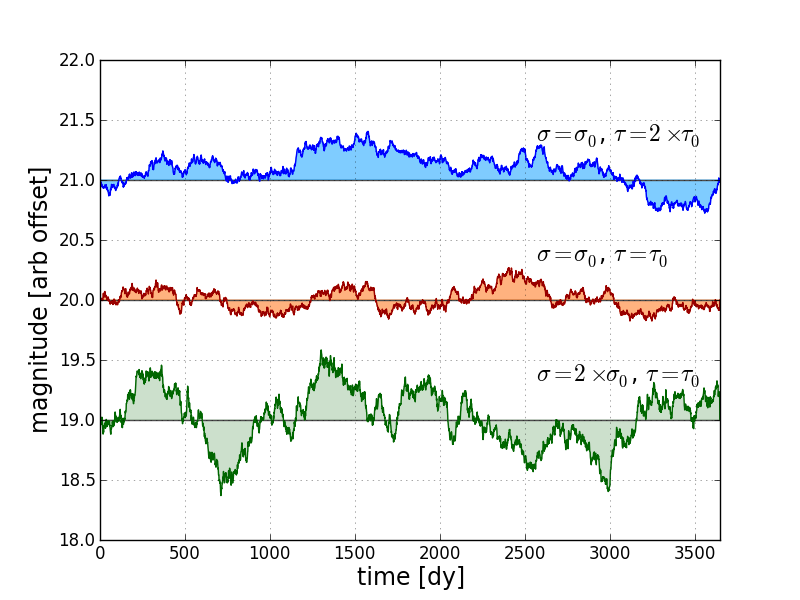}
\caption{\label{fig:example_lcs}. Examples of quasar light curves generated by the CAR model, with different variability amplitude $\sigma$ and characteristic time-scale $\tau$ ($\sigma_0=0.01$ mag day$^{-1/2}$ and $\tau_0=300$ day in this example). }
\end{center}
\end{figure*}

The optical light curves of quasars arise from fluctuations in the
brightness of the accretion disk with structure in the time series on
the order of days.  Since these fluctuations are coherent, the
implication is that the size of the accretion disk is roughly $R_{\rm
src} \sim 10^{16}$ cm (which will be important for the microlensing
calculation in Section~\ref{sec:microlensing}).  These fluctuations
have been found to be well described by a damped random walk (DRW)
stochastic process \citep[see e.g.\
][]{KellyEtal2009,MacLeodEtal2010,ZuEtal2013}.  Initially,
\citet{KellyEtal2009} introduced the Continuous Auto Regressive (CAR)
process for fitting quasar light curves; this is equivalent to a
Gaussian Process in which the covariance between two points on the
light curve decreases as a function of their temporal separation. The
CAR process is given by \citep[see the Appendix in][]{KellyEtal2009},
\begin{equation}
  M(t) = e^{-t/\tau} M(0) + \bar{M}(1-e^{-t/\tau}) + \sigma\int_{0}^{t} e^{-(t-s)/\tau} dB(s),
\end{equation}
where $M$ is the magnitude of an image, $\tau$ is a characteristic
timescale in days, $\bar{M}$ is the mean magnitude of the light curve in
the absence of fluctuations, and $\sigma$ is the characteristic
amplitude of the fluctuations in mag/day$^{1/2}$.  In this model,
fluctuations are generated by the integral term where $dB(s)$ is a
normally distributed value with mean zero and variance $dt$.
By fitting the above model to some 100 MACHO project light curves,
\citet{KellyEtal2009} generated a distribution of $\tau$ and $\sigma$
for the MACHO quasars; we show typical examples of the CAR process with
reasonable values for those parameters in Figure~\ref{fig:example_lcs}.
Likewise, \citet{MacLeodEtal2010} fit the DRW model to over 9000
quasars in the SDSS Stripe 82 region, again finding it to be a good
description of the data, and exploring correlations between $\tau$ and
$\sigma$, and quasar luminosity and black hole mass.

While the DRW process provides a good description of the
data obtained so far, it is not yet clear whether it will remain so for
longer baseline, higher cadence, or multi-filter light curves. The
different emission regions of an AGN (different parts of the accretion
disk, broad and narrow line clouds, etc.) are likely to vary in
different ways \citep{EigenbrodEtal2008a,SluseEtal2011,SluseEtal2012},
suggesting that linear combinations of stochastic processes could
provide more accurate descriptions \citep{KellyEtal2011}.  These
subcomponents would likely need parameters drawn from different
distributions to the one above, and the correlations between the
processes may need to be taken into account as well.  Nevertheless, the
success of the CAR model to date makes it a reasonable place to begin
when simulating LSST-like AGN light curves.


\subsection{Multiple Imaging by a Foreground Galaxy}
\label{sec:time_delay_dist}

For a given lens system, the time delays between images can be as short
as $\sim$1 day for close pairs of images to as long as $\sim$100s of
days for images on opposite sides of the lensing galaxy.  The magnitude
of these time delays (as well as the other observables) depends on the
redshifts of both the lens galaxy, $z_l$, and the source redshift, $z_s$,
and therefore it is important to understand the expected distribution of
those parameters in the LSST sample.  \citet[][hereafter OM10]{OM10}
generated a mock catalog of LSST lensed AGN based on plausible models
for the source quasars and lens galaxies, and simple assumptions for the
detectability of lensed quasars, including published 10$\sigma$ limiting
magnitude estimates, and the assumption that lenses will be detected if
the third (second) brightest image for a given quad (double) is above
this limit. This catalog provides a distribution of time delays that
will be present in the LSST data which we can use to guide the generation of
mock light curves.

\begin{figure*}[!ht]
\begin{center}
\includegraphics[width=0.9\textwidth]{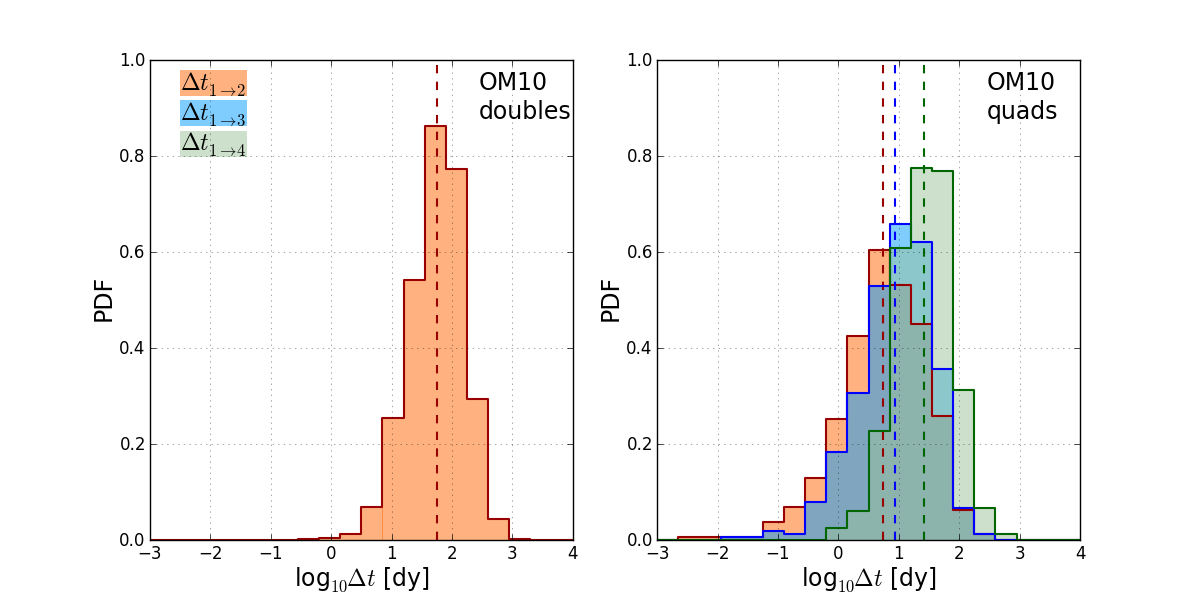}
\caption{\label{fig:OM10dt} The distribution of expected time delays in the OM10 mock lens catalog. Left: doubles, Right: quads. Dashed lines mark the means of the $\log_{10} \Delta t$ distributions.}
\end{center}
\end{figure*}

Figure~\ref{fig:OM10dt} shows the $\log_{10} \Delta t$ distributions for
the OM10 double and quad sample.  The distributions are roughly
log-normal with means $\sim$10s of days and tails extending below 1 day
for the quads, and above 100 days for the doubles. Lenses in both of
these tails will have time delays that are difficult to measure, either
because the cadence isn't high enough, or because the observing seasons
are not long enough. We expect some fraction of time delay measurements
to fail catastrophically in these cases, but we also expect the
catastrophe rate (and the robustness with which failure is reported) to
vary with measurement algorithm.

\subsection{Microlensing}
\label{sec:microlensing}

As noted in Section~\ref{sec:car}, the physical size of a quasar
accretion disk is $R_{\rm src} \sim 10^{15}$-$10^{16}$ cm, which, at
cosmological distances, represents an angular size of $\sim1$
$\mu$arcsecond ($\mu$as).  In addition, the Einstein radius for a 1
$M_{\odot}$ point mass at these distances is also $\sim1$ $\mu$as,
indicating that the stars in the lens galaxy will typically have an
order unity (or more) effect on the brightnesses of the individual
images.  Given the relevant angular scales, this phenomenon is termed
``microlensing''.

Microlensing has long been acknowledged as a significant source of
potential error when estimating time delays from optical monitoring data
\citep[see e.g.][and references therein]{Schild,SchechterEtal1997,TewesEtal2013a} due to the fact
that the relative velocity between the source and lens leads to time
dependent fluctuations that are independent between the images.
For caustic crossing events the relevant time scales are months to
years, with smoother variations occurring over roughly decade
timescales.  As expected, the microlensing fluctuations are larger at
bluer wavelengths, which correspond to smaller source sizes
\citep[e.g.,][]{Kochanek2004,MorganEtal2008b}. A solution to
measuring time delays in the presence of these fluctuations (which are
uncorrelated between the quasar images) is to model the microlensing in
each image individually at the same time as inferring the time delay
\citep[e.g.][]{Kochanek2004,TewesEtal2013b}.

We create mock microlensing signals in each quasar image light curve by
calculating the magnification as the source moves behind a static
stellar field.  The parameters involved are the local convergence,
$\kappa$, and shear, $\gamma$, the fraction of surface density in stars,
$F_{\star}$, the source size, $R_{\rm src}$,  and the relative velocity
between the quasar and the lens galaxy, $v_{\rm rel}$.  We also include a
Salpeter mass function for the stars though the amplitude of the
fluctuations depends predominantly on the mean mass (which we take to be
1 $M_{\odot}$).\footnote{The microlensing code used in this work, {\sc
MULES} is freely available at
\texttt{https://github.com/gdobler/mules}.}

For each lens in the OM10 catalog we assign an $F_{\star}$ at each
image position as follows.  The OM10 catalog provides the velocity
dispersion for a given lens which we use to estimate the $i$-band
luminosity and effective radius of the galaxy by drawing from the
Fundamental Plane \citep[e.g.\ ][]{BernardiEtal2003}.  For a given
mass-to-light ratio, and assuming a standard \citet{deVaucouleurs1948}
profile for the brightness distribution centered on the lens with an
isothermal ellipsoid for the total mass distribution, we obtain the
ratio of stellar mass density to total mass density, $F_{\star}$, at
each image position.

\begin{figure}[!ht]
\begin{center}
\includegraphics[width=0.9\columnwidth]{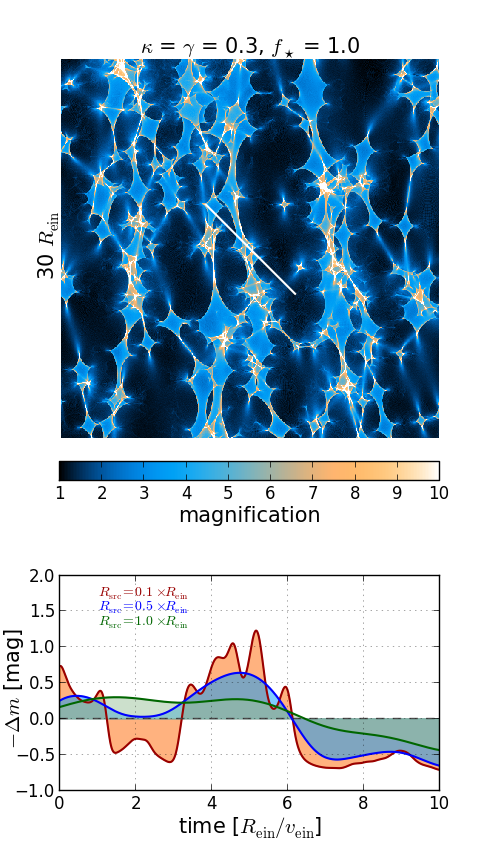}
\caption{\label{fig:ml}
\emph{Top}: A typical source plane magnification map, showing the
complex caustic structures caused by the stars in the lens galaxy. As
the background quasar and foreground galaxy move relative to each
other, the source traverses this pattern (white solid line), resulting
in brightness fluctuations due to microlensing.  \emph{Bottom}: The
fluctuations along this track (which traverses 10 stellar Einstein
radii) are shown for three representative source sizes.  Larger source
sizes decrease the amplitude of the microlensing fluctuations as well
as smooth out the shorter time scale features.  Here ``time'' is given
in units of Einstein radii crossing time (e.g., if the relative
velocity between the lens and source is 1 $R_{\rm ein}$/yr [see text],
then this track represents a 10 yr light curve.)}
\end{center}
\end{figure}

Given $\kappa$ and $\gamma$ from the OM10 catalog and estimating
$F_{\star}$ as above, we generate magnification maps like the one
shown in Figure~\ref{fig:ml}, which represent the magnification of a
point source as a function of position in the source plane.  To use
this map to generate temporal microlensing fluctuations, we first
smooth it by a Gaussian source profile of appropriate size ($\sim
10^{15}$-$10^{16}$ cm) and then trace a linear path along a random
direction in the map.  This path is converted from source plane
position to time units via a relative velocity $v_{\rm rel}$
\citep{KayserEtal1986} which we compute from the velocities of
matching galaxies drawn from the Millennium
Survey.\footnote{\url{http://gavo.mpa-garching.mpg.de/Millennium/}}
The effect of having a finite source is to smooth out and reduce the
amplitude of the microlensing fluctuations.


\subsection{Sampling}
\label{sec:sampling}

The current state-of-the-art lens monitoring campaign, COSMOGRAIL,
typically visits each of its targets every few nights during each of
several observing seasons each lasting many months. For example,
\citet{TewesEtal2013b} present 9 seasons of monitoring for the lensed
quasar RXJ1131$-$1231 where the mean season length was 7.7 months ($\pm
2$ weeks) and the median cadence was 3 days. These observations were
taken in the same R-band filter, with considerable attention paid to
photometric calibration and PSF estimation based on the surrounding star
field. These data allowed \citet{TewesEtal2013b} to measure a time delay
of 91 days to 1.5\% precision.

\begin{figure*}[!ht]
\begin{center}
\includegraphics[width=0.9\textwidth]{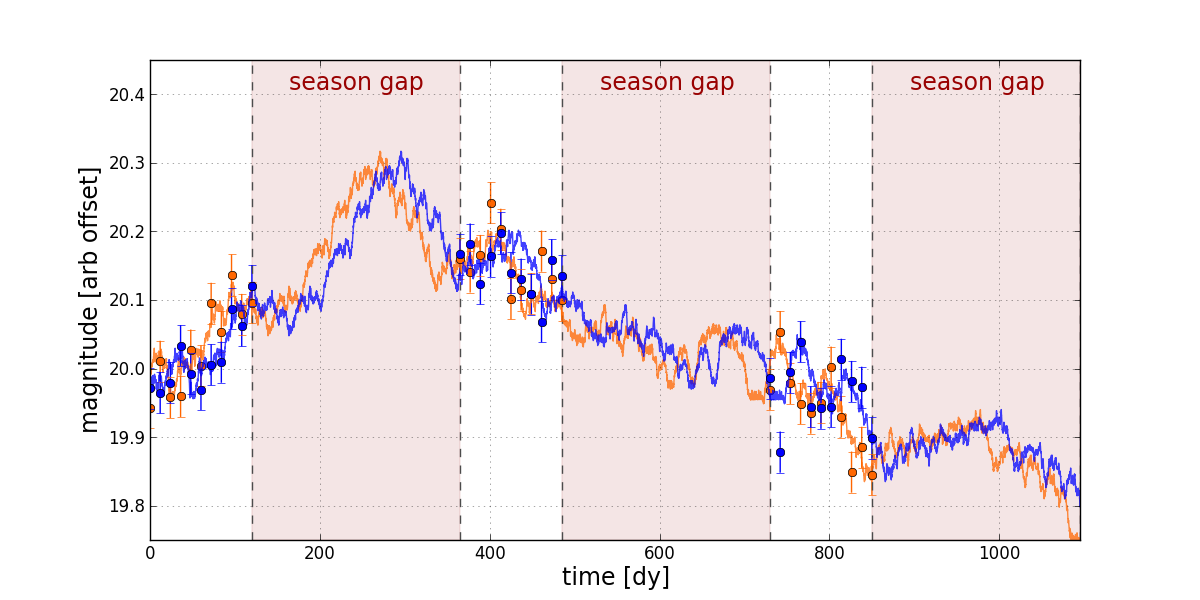}
\caption{\label{fig:sample}
Example light curves for a simulated double lensed quasar. The blue
light curve lags behind the red light curve as a result of the
gravitational time delay.  The filled circles with error bars
represent an actual mock observation in which noise and measurement
uncertainty are added, while the finite season lengths lead to gaps in
the data.}
\end{center}
\end{figure*}

While this quality of measurement is possible for small samples (a few
tens) of lenses, the larger sample of lensed quasars lying in the LSST
survey footprint will all be monitored over the course of its ten year
campaign, but at lower cadence and with shorter seasons. In the simplest
possible ``universal cadence'' observing strategy, we would expect the
mean cadence to be around 4 days between visits, in any filter, and with
some variation with time as the scheduler responds to the needs of the
various science programs and the changing conditions; the gaps between
observations in the same filter will tend to be longer
\citep{LSSTpaper,LSSTSciBook}. The season length in this strategy is
likely to be approximately 4 months (with variation among filters), in
order to keep the telescope pointing at low airmass (see example in
Figure~\ref{fig:sample}). The primary impact of the shorter season
length will be to make it hard to measure time delays of more than 100
days; the LSST universal cadence time delay lens sample would be biased
towards delays shorter than this.

The universal cadence strategy may not turn out to be optimal, and we
can explore various LSST observing strategies by simulating light curves
with a range of cadences and season lengths. Shorter cadences and
longer seasons are closer to those obtained by COSMOGRAIL. As its lens sample increases in size, blind analysis of the COSMOGRAIL datasets will provide an increasingly better understanding of the accuracy available to the program. We note that
only if all filters' light curves can be fitted simultaneously with a
model for the multi-filter variability would the maximum, any-filter
cadence be fully exploited. Even if that fitting is not possible, the
dithered nature of the different filters' light curves should still
allow a time resolution approaching that of the any-filter
cadence.

The remaining variables in the mock light curve generation pertain to the
photometric uncertainties applied to the sampled fluxes.
\citet{TewesEtal2013a} provide a summary of possible sources of
uncertainty and error in the photometric measurements, which we follow
in generating light curves with realistic uncertainties, including in
the accuracy of the error reporting. The OM10 mock lens sample
contains a variety of quasar image brightnesses, allowing us to
investigate time delay accuracy as a function of signal to noise, or,
for LSST, source magnitude. {\bf We note that for this first
challenge, uncertainties arising from contamination by the light of
the foreground source were not taken into account. Those might be
important, especially for the fainter images and this should be
addressed in future challenges.}

\section{The Structure of the Challenge}
\label{sec:structure}

This section outlines the two initial steps of the challenge, gives the
instructions for participation and timeline, and defines the goal of the
challenge and the criteria for evaluation.


\subsection{The Challenge Ladders}
\label{ssec:steps}

The initial challenge consists of two parts, hereafter time delay
challenge 0 and 1 (TDC0 and TDC1). Each of these is organized as a
ladder with a number of simulated light curves at each rung. The rungs
are intended to represent increasing levels of difficulty and realism
within each challenge.  The simulated light curves were created by the
``evil'' team (authors GD, CDF, KL, PJM, NR, and TT). All the details
about the light curves, including input parameters, noise properties,
etc., were only revealed to the participating teams
(hereafter ``good'' teams) after the closing of the challenge\footnote{We note here that the tongue-in-cheek names ``evil'' and ``good'' teams do not denote any despicable intention or moral judgment, but were chosen to capture the desire of the challenge designers to produce significantly realistic (and difficult) light curves as well as an incentive for the outside teams to participate.}.

TDC0 consists of a small number of simulated light curves with fairly
basic properties in terms of noise, sampling season, and cadence. It
is intended to serve as a validation tool before embarking on
TDC1. The ``evil'' team expects that state of the art algorithms should be
able to process TDC0 with minimal computing time and recover the input
time delays within the estimated uncertainties. TDC0 also provides a
means to perform basic debugging, and to test input and output formats
for the challenge. The truth file for TDC0 will not be revealed until
after the closing of TDC1 to preserve blindness.

TDC1 consists of thousands of sets of simulated light curves, also
arranged in rungs of increasing difficulty and realism. The large data
volume is chosen to simulate the demands of an LSST-like experiment, and
to be able to detect biases in the algorithms at the sub-percent level.
The ``evil'' team expects processing of the TDC1 dataset to be
challenging with current algorithms in terms of computing resources.
TDC1 thus represents a test of the accuracy of the algorithms but also
of their efficiency. Incomplete submissions were accepted, although
the number of processed light curves is one of the metrics by which
algorithms were evaluated, as described below.

The mock data generated for the highest rungs of the initial challenge
ladders TDC0 and TDC1 are as realistic as our current simulation
technology allows, but lower rungs are somewhat simplified. This design
is based on the successful weak lensing STEP \citep{STEP1,STEP2} and
GREAT \citep{GREAT08,GREAT10Stars,GREAT10Galaxies,GREAT3} shape
estimation challenges, where the former tried to be as realistic as
possible, while the latter focused on specific aspects of the problem.
Still, following a successful outcome of TDC0 and TDC1 we anticipate in
the future further increasing the complexity of the simulations so as to
stimulate gradual improvements in the algorithms over the remainder of
this decade.  Our approach of testing on simulated data is
very complementary to tests on real data. The former allow one to test
blindly for accuracy but they are valid only insofar as the simulations
are realistic, while the latter provide a valuable test of consistency
(but not accuracy) on actual data, including all the unknown unknowns.


\subsection{Instructions for participation, timeline, and ranking criteria}
\label{ssec:instruction}

Instructions for how to access the simulated light curves in the time
delay challenge are given at the challenge
website\footnote{\url{http://timedelaychallenge.org}}.
In short, participation in the challenge required the following steps.

\subsubsection{TDC0}

Every prospective ``good'' team was invited to download the TDC0 light curves
and analyze them. Upon completion of the analysis, they were asked to submit
their time delay estimates, together with their estimated 68\%
uncertainties, to the challenge organizers for analysis. The simulation
team calculated a minimum of four standard metrics given this set of
estimated time delays $\tilde{\Delta t}$ and uncertainties $\delta$.
The first metric is efficiency, quantified as the fraction of light curves
$f$ for which an estimate is obtained.  Of course, this is not a
sufficient requirement for success, as the estimate should also be
accurate and have correctly estimated uncertainties. There might be instances where the data are ambiguous (e.g., if time delay falls into season
gaps), in which case some methods will indicate failure while others will
estimate very large uncertainties.

Therefore, we need to introduce a second metric to evaluate how realistic
the error estimate is. For this, we use the
goodness of fit of the estimates, quantified by the standard reduced
$\chi^2$:
\begin{equation}
\chi^2=\frac{1}{fN}\sum_i \left(\frac{\tilde{\Delta t}_i - \Delta t_i}{\delta_i}\right)^2.
\end{equation}
where $\Delta t_i$ are the true time delays defined positive in input.

The third metric is the claimed precision of the estimator, quantified
as the average relative uncertainty per lens:
\begin{equation}
P=\frac{1}{fN}\sum_i \frac{\delta_i}{\Delta t_i}.
\end{equation}

The fourth is the accuracy of the estimator, quantified by the average
fractional residual per lens
\begin{equation}
A=\frac{1}{fN} \sum_i \frac{\tilde{\Delta t}_i - \Delta t_i}{\Delta t_i}.
\end{equation}

The initial function of these metrics is to define a minimal
performance threshold that must be passed, in order to guarantee
meaningful results in TDC1. ``Good'' teams are given aggregated
statistical feedback on their TDC0 efforts, from which they can decide
whether to continue to TDC1.  The criteria for passing the TDC0 test
are as follows:

\begin{eqnarray}
f>0.3    \\
0.5<\chi^2<2  \\
P<15\%   \\
|A|<15\%
\end{eqnarray}

A failure rate of 70\% is something like the borderline of acceptability
for LSST (given the total number of lenses expected), and so can be used
to define the efficiency threshold. The TDC0 lenses were selected to
span the range of possible time delays, rather than being sampled from
the OM10 distribution, and so we therefore expect a higher rate of
catastrophic failure at this stage than in TDC1: 30\% successes is a
minimal bar to clear.

The factor of two half-ranges in reduced $\chi^2$ correspond
approximately to fits that include approximately 95\% of the $\chi^2$
probability distribution when $N=8$, i.e. the number of time delays in
every rung of TDC0: fits outside this range likely have problems with
the time delay estimates, or the estimation of their uncertainties, or
both. Requiring an average precision and accuracy of better than 15\%
is a further minimal bar to clear.

Repeat submissions were accepted for teams to iterate their
analyses. TDC0 remained blinded until the TDC1 deadline on 1 July
2014. Late TDC0 submissions were accepted, but those teams had less
time to carry out TDC1.
\begin{table*}
\begin{center}
\begin{tabular}{cccccc} \hline\hline
  Rung &  Sampling &  Season duration   &  Noise  &   Microlensing   \\ \hline
  0    &  1 dy        &    12 mon                    &   0.03 uni       &  no          \\
  1    &  1 dy        &    4 mon                    &   0.03 uni       &   no           \\
  2    &  1 dy        &    4 mon                    &   opsimish      &   no              \\
  3    &  2 wk        &    12 mon                    &   0.03 uni       &   no             \\
  4    &  2 wk       &    4 mon                    &   0.03 uni       &   no             \\
  5    &  opsimish      &    4 mon                    &   opsimish       &  no          \\
  6    &  opsimish       &    4 mon                    &   opsimish       &   yes            \\

\hline\hline
\end{tabular}
\caption{Design for 7 rungs in TDC0. The opsimish for sampling is a Gaussian distribution with mean sampling 12 days and deviation 2 days while
   the opsimish for noise is  0.053 in nanomaggies with error 0.016}
\end{center}
\end{table*}

\begin{figure*}[!htbp]
\begin{center}
\includegraphics[width=175mm]{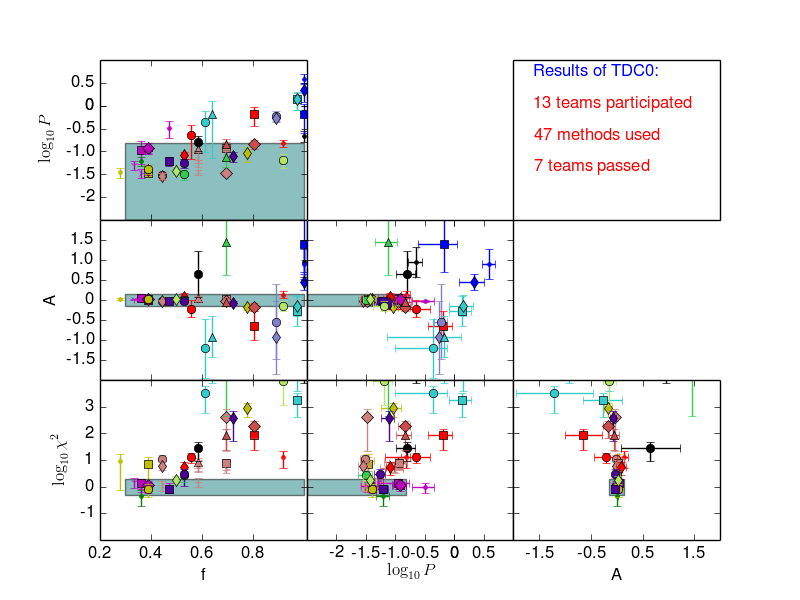}
\caption{Results of TDC0. Each color represents a different team, while each symbol represents a different method. The teams and methods are not identified to preserve confidentiality of the TDC0 submission.
 }\label{fig:TDC0}
\end{center}
\end{figure*}

As of July 1 2014, the closing date of TDC1, 13 teams participated in
TDC0, using 47 different algorithms. Of those teams, seven qualified for
TDC1. A summary of the results is shown in Figure~\ref{fig:TDC0}. The
seven qualified teams are revealed in a companion paper, where their
TDC1 submissions are analyzed.

\subsubsection{TDC1}

``Good'' teams that successfully passed TDC0 and wished to continue
were given access to the full TDC1. As in TDC0, the ``good'' teams
estimated time delays and uncertainties and provided the answers to
the ``evil'' team via a suitable web interface (found at the challenge
website). The ``evil'' team computed the metrics described above.  The
results were not revealed until the end of the challenge in order to
maintain blindness.

The deadline for TDC1 was 1 July 2014, seven months after that of
TDC0.  Multiple submissions were accepted from each team in order to
allow for correction of bugs, and for different algorithms.  However,
only the first submission was considered blind. The most recent
submission for each algorithm was also considered in order to allow
for teams to improve their methods. Late submissions were accepted and
included in the final publication if received in time but were
flagged as such.

\subsubsection{Publication of the results}

Initially this first paper was only posted on the arXiv as a means to
open the challenge.  After the TDC1 deadline, this paper has been
revised to include the details and results of TDC0.  At the same time,
the full details and results of TDC1 are described in the second paper
of this series, including as co-authors all the members of the
``good'' teams who participated in the challenge. The two papers were
submitted concurrently so as to allow the referee to evaluate the
entire process. ``Good'' teams have been encouraged to publish papers
on their own methods making use of the challenge data, if they felt
they are presenting innovation worthy of publication.


\subsection{Overall goals and broad criteria for success}

The overall goal of TDC0 and TDC1 is to carry out a blind test of
current state of the art time delay estimation algorithms in order to
quantify the available accuracy. Criteria for success depend on the
time horizon. At present, time delay cosmology is limited by the
number of lenses with measured light curves and by the modeling
uncertainties which are of order 5\% per system
\citep[e.g.,][]{SuyuEtal2010,SuyuEtal2013}. Furthermore, distance
measurements are currently in the range of accuracy of 3\%. Therefore,
any method that can currently provide time delays with realistic
uncertainties ($\chi^2<1.5$) for the majority ($f>0.5$) of light curves
with accuracy $A$ and precision $P$ better than 3\% can be considered a
competitive method.

In the longer run, with LSST in mind, a desirable goal is to maintain
precision of $P<3\%$ per lens, but to improve the accuracy to $|A| <
0.2\%$ in order for the sub-percent precision cosmological parameter
estimates not to be limited by time delay measurement systematic
biases.  For $N=1000$, the 95\% goodness of fit requirement becomes
$\chi^2 < 1.09 fN$, while keeping $f>0.5$. Testing for such extreme
accuracy requires a large sample of lenses: TDC1 contains several
thousand simulated systems to enable such tests.


\section{Summary}
\label{sec:summary}

Strong lens time delays are a powerful tool for cosmology. Like every
cosmographic probe, in order to reach the precision and accuracy
necessary to measure the dark energy equation of state it is essential
to subject every component of the method to rigorous testing. With
this motivation we have initiated a time delay challenge (TDC). In
this paper, we have described the tools developed and used by the
``evil team'' to construct simulated light curves, laid out the
structure of the challenge to the community, and given the results of
TDC0.

The intrinsic quasar light curves are generated using a damped random
walk process. The multiple images, flux ratios, and time delays are taken
from the properties of a realistic simulated catalog of lenses expected
for a survey like LSST. The effects of microlensing are computed using a
newly developed fast code. Realistic noise and monitoring patterns are
applied to the data. All simulation software is written in python and
will be made publicly available after the completion of the challenge.

The challenge consists of two steps, TDC0, consisting of a few pairs
of image light curves, intended to provide ``good'' teams with the
opportunity to test and debug their codes before launching into the
more computationally intensive TDC1.  In total, 13 teams participated
in TDC0 using 47 different methods. Seven of those teams qualified for
TDC1.  The TDC1 dataset consists of thousands of light curves, a
number sufficient to identify biases at the subpercent level required
for Stage IV experiments.

The challenge data are available to download at

\begin{center}
\url{http://timedelaychallenge.org}
\end{center}

The deadlines for the challenges were December 1, 2013 for TDC0, and
July 1, 2014 for TDC1. The challenge is (still) open to anyone. The
results of TDC1 are published in a companion paper with all the
participating teams as co-authors \citep{LiaoEtal2014}.


\acknowledgments

We thank Frederic Courbin, Malte Tewes and Brendon Brewer for useful
comments and suggestions about the challenge. We acknowledge the LSST
Dark Energy Science Collaboration for hosting several meetings of the
``evil'' team, and the private code repository used in this work.  CDF
and TT acknowledge support from the NSF through Collaborative Award
``Accurate cosmology with strong gravitational lens time delays'',
(AST-1312329 and 1450141). TT acknowledges support from the Packard
Foundation through a Packard Research Fellowship. PJM and EL
acknowledge the U.S.\ Department of Energy Office of Science under
Contracts No.\ DE-AC02-76SF00515 and DE-AC02-05CH11231
respectively. KL was supported by the China Scholarship Council. AH is
supported by an NSERC discovery grant. This paper was drafted using
the Authorea web service at
\url{http://authorea.com}.


\bibliographystyle{apj}
\bibliography{references}


\end{document}